\documentclass[letterpaper]{article}
\usepackage{aaai}
\usepackage{times}
\usepackage{helvet}
\usepackage{courier}
\usepackage{amsmath}
\usepackage{amssymb}
\usepackage{graphicx}
\usepackage{booktabs}
\usepackage{url}
\usepackage{multirow}
\usepackage{makecell}
\nocopyright
\title{Training-Free Prompt-Driven Audio-Visual Editing}
\author{Shaohua Pan\textsuperscript{}\thanks{Equal contribution.} \and Xiao Liu\textsuperscript{}\footnotemark[1] \\
}

\begin{document}

\title{AV-FlowEdit: Training-Free Prompt-Driven Audio-Visual Editing with Event-Aware Spectral Trajectory Preservation}
\maketitle

\begin{figure*}[t]
    \centering
    \includegraphics[width=\textwidth]{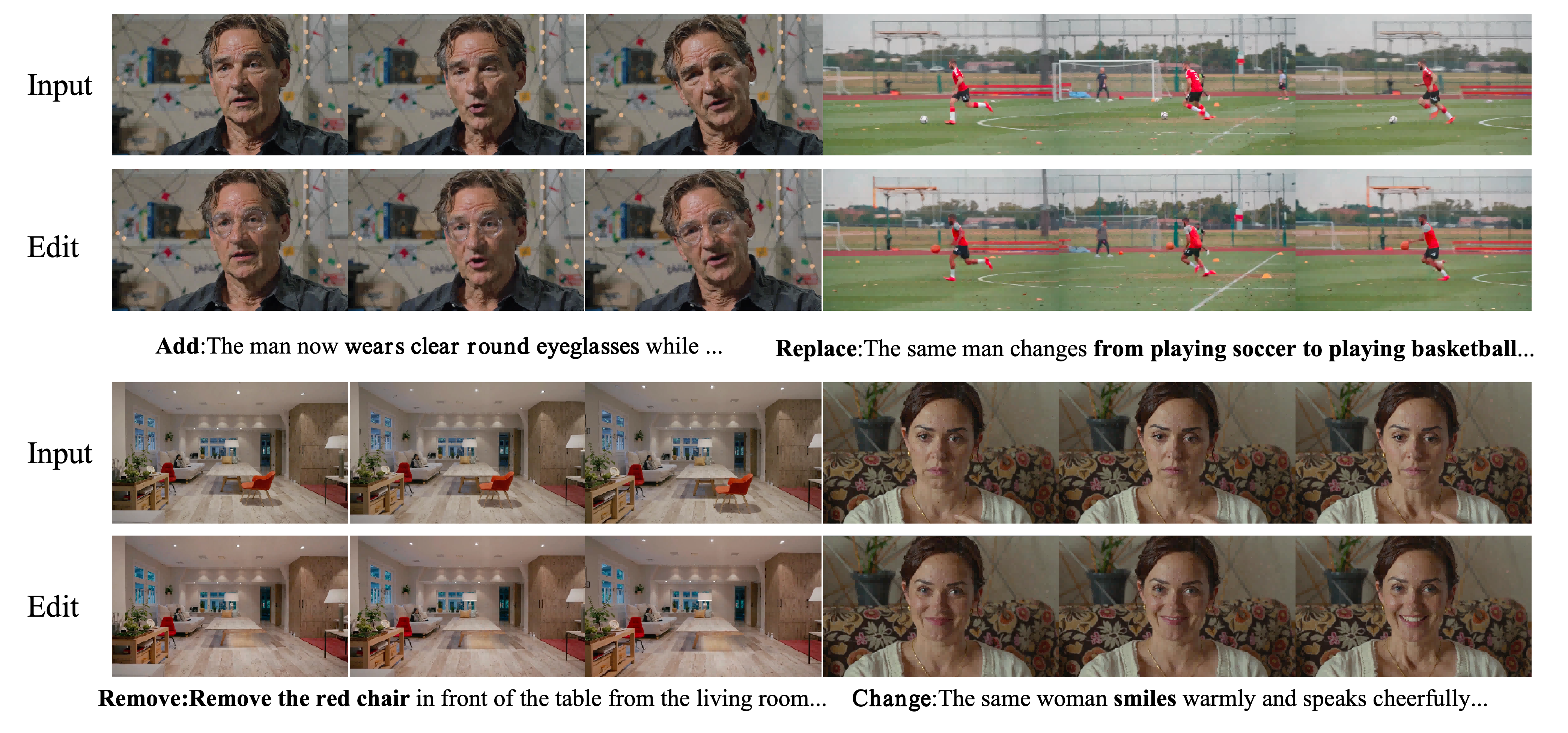}
    \caption{Results of our training-free audio-video editing framework. Given a source video and a target prompt, our method performs consistent semantic editing while preserving scene structure, camera motion, and audio-video synchronization.}
    \label{fig:intro}
\end{figure*}
\begin{abstract}
Joint audio-video diffusion models generate synchronized content, but editing existing clips requires target changes without disrupting non-target structure or cross-modal correspondence. We present AV-FlowEdit, a training-free and inversion-free framework for prompt-driven audio-video editing. Our key insight is that source--target flow differences provide both cross-modal editing directions and preservation cues. Under shared perturbations, they jointly update the video and audio branches, supporting visual-only, audio-only, and joint edits while retaining pretrained cross-modal interactions. To suppress drift accumulated during sampling, Cross-Modal Event-Aware Spectral Trajectory Preservation combines flow saliency, aligned audio activity, multiscale low-frequency displacement, and diffusion-stage scheduling. This adaptively preserves unsupported regions while allowing intended audio-visual events. AV-FlowEdit requires no model training, external masks, or optimization-based inversion. Experiments show strong prompt alignment and temporal consistency, together with competitive audio quality and cross-modal coherence.
\end{abstract}

\section{Introduction}
Recent advances in video generative modeling have substantially improved spatial fidelity, temporal coherence, and text alignment~\cite{ho2022video,lipman2022flow,peebles2023scalable,kondratyuk2024videopoet}. Joint audio-video generative models further extend this progress by jointly modeling visual content, audio, and text within a unified generation process, enabling them to capture cross-modal event semantics and temporal correspondence~\cite{ruan2023mm,kim2024versatile,wang2024av,ishii2025simple}. Editing an observed clip, however, imposes stricter constraints than generating a new sample: the model must introduce the requested change while preserving source identity, background layout, camera motion, and audio-visual correspondence outside the intended edit~\cite{ku2024anyv2v,zhang2025v2edit,mai2025easyv2v}.

Existing video editing methods can be broadly organized by whether they require model adaptation and, among training-free approaches, whether they reconstruct a source diffusion trajectory. Model-adaptation methods optimize pretrained generators for a source video or editing task, but introduce per-instance or task-specific training costs~\cite{wu2023tune,shin2024edit}. Training-free methods avoid parameter updates and instead modify the inference process. Many such methods rely on diffusion inversion, first recovering a source noise or latent trajectory and then resampling it under a target prompt; however, inversion errors can accumulate over time, while strong source-feature constraints may restrict large changes in content, action, or scene~\cite{qi2023fatezero,liu2023video,ceylan2023pix2video,geyer2024tokenflow}. More recently, inversion-free flow-based methods estimate editing directions directly from differences between source- and target-conditioned velocity fields, avoiding explicit reconstruction of the source trajectory~\cite{kulikov2025flowedit,kim2025flowalign}. Related flow-steering principles have also been extended to video editing, providing a lightweight route to editing video content, actions, and dynamics without per-example inversion or training~\cite{li2025flowdirector,kulikov2026versatile}.

Despite these advances, extending inversion-free flow-based editing to general audio-video editing presents two unresolved challenges. First, existing methods operate primarily on the visual modality and do not jointly update the accompanying audio. Pairing them with a separate audio editing system bypasses the bidirectional interactions learned by a joint audio-video model, potentially weakening the semantic and temporal correspondence between the two modalities. Second, iterative source-target flow updates can accumulate unintended deviations across sampling steps, leading to background drift, subject deformation, and inconsistent camera motion. Simple preservation mechanisms introduce their own trade-offs: fixed source-feature injection may suppress the requested edit, spatial masks depend on reliable region-level priors, and uniform low-frequency constraints may remove the coarse spatiotemporal changes required by intended actions or synchronized audio-visual events.

To address both challenges, we propose AV-FlowEdit, a training-free framework that jointly edits audio and video using a frozen joint audio-video generative model. At each sampling step, AV-FlowEdit uses differences between source- and target-conditioned velocity fields to update both modalities, thereby retaining their cross-modal interactions without explicit source inversion. Our key idea is to use these source--target flow differences not only to drive the desired edit, but also to identify where source structure should be preserved. Strong flow responses indicate spatiotemporal positions where target-driven changes are supported, whereas weak responses identify content that should remain close to the source. To suppress accumulated structural drift, we decompose the source-to-edit video displacement across multiple spatiotemporal scales and adaptively correct unsupported components according to flow saliency, target-audio activity, and the diffusion stage. This correction preserves non-target structure while allowing the coarse spatiotemporal changes required by intended audio-visual events. Fig.~\ref{fig:intro} presents representative examples of target-driven editing and non-target scene preservation with AV-FlowEdit. In conclusion, our main contributions include:

\begin{itemize}

\item We propose AV-FlowEdit, a training-free and inversion-free framework for prompt-driven audio-video editing with pretrained joint generative models. By computing source--target difference flows for video and audio within the same frozen transformer, AV-FlowEdit supports visual-only, audio-only, and joint edits in a unified inference procedure while retaining the pretrained model's cross-modal interactions. It requires neither per-example inversion nor separate modality-specific editing pipelines.

\item We introduce Cross-Modal Event-Aware Spectral Trajectory Preservation, which uses source--target flow saliency and temporally aligned target-audio activity to identify unsupported changes, decomposes accumulated video displacement into multiscale low-frequency structural components, and schedules their correction according to the diffusion noise level. This adaptive mechanism suppresses non-target structural drift while allowing the coarse spatiotemporal changes required by target-supported audio-visual events.

\item We evaluate AV-FlowEdit against independent, cascaded, and joint audio-video editing baselines using metrics for target alignment, source preservation, temporal stability, audio quality, and cross-modal coherence. The results show strong target alignment and temporal consistency, together with competitive audio and cross-modal performance, under a training-free and inversion-free setting.

\end{itemize}

\section{Related Work}

\subsection{Video and Audio-Video Foundation Models}

Video generation has evolved from spatiotemporal diffusion models, such as Lumiere~\cite{bar2024lumiere}, to large-scale DiT-based architectures, including Wan~\cite{wan2025wan}, HunyuanVideo~\cite{kong2024hunyuanvideo}, and LTX-Video~\cite{hacohen2024ltx}. Recent models typically adopt diffusion, rectified flow, or flow matching objectives, achieving substantial improvements in visual fidelity and temporal consistency~\cite{ho2022video,lipman2022flow,esser2024scaling}. However, these models are primarily designed for content generation from noise rather than structure-preserving editing of observed videos.

Most existing systems synthesize video and audio separately, making event-level semantic and temporal synchronization difficult\cite{hua2025vabench,ren2025sta,dai2026omni2sound}. LTX-2~\cite{hacohen2026ltx} instead employs separate video and audio branches that exchange information through cross-attention, enabling joint audio-video generation while preserving modality-specific representations. Nevertheless, its generation and retake pipelines are not designed for source-to-target prompt editing, which requires target-driven changes and source-content preservation to be balanced within the same inference process.

\subsection{Training-Based Video Editing}

Training-based video editing adapts pretrained generative models to a source video or a specific editing task. One-shot methods such as Tune-A-Video and Edit-A-Video optimize model parameters using a single source video to preserve its appearance and motion while enabling prompt-guided modifications~\cite{wu2023tune,shin2024edit}. Although these methods reduce the need for large paired editing datasets, they still require source-specific optimization and may incur additional storage and computational costs for each input video. Training-free methods instead keep the pretrained model parameters fixed and control the editing process through inference-time operations, avoiding per-video model adaptation.

\subsection{Training-Free Video Editing}

Training-free video editing can be broadly divided into inversion-based and inversion-free methods.

\textbf{Inversion-based editing.}
Early training-free approaches are largely built upon diffusion inversion, which reconstructs the latent noise or diffusion trajectory of a source video before resampling under a target prompt. Representative methods include FateZero~\cite{qi2023fatezero}, Video-P2P~\cite{liu2023video}, Pix2Video~\cite{ceylan2023pix2video}, and TokenFlow~\cite{geyer2024tokenflow}. These methods combine source inversion with attention control, feature injection, or feature propagation to preserve source appearance and temporal consistency during editing. However, inversion errors can accumulate over long video sequences, while strong source constraints may limit large motion and semantic edits. Extending inversion to joint audio-video generation is further complicated by heterogeneous latent spaces and temporal resolutions across modalities.

\textbf{Inversion-free editing.}
Recent methods avoid explicit inversion by directly modifying the generation trajectory. FlowEdit~\cite{kulikov2025flowedit} edits videos using the difference between source- and target-conditioned flow fields, while FlowAlign~\cite{kim2025flowalign}, FlowDirector~\cite{li2025flowdirector}, and DynaEdit~\cite{kulikov2026versatile} further improve trajectory consistency, region-aware control, and dynamic content editing. More recently, OmniEdit~\cite{lin2026omniedit} extends training-free editing to the audio-visual domain by jointly enabling lip synchronization and content editing without model fine-tuning. 

Despite these advances, existing methods either focus on visual editing or target specialized audio-visual scenarios such as talking-head lip synchronization, rather than general-purpose audio-video semantic editing. Moreover, their preservation strategies typically rely on fixed attention guidance or low-frequency constraints, making it difficult to adaptively balance structural preservation and semantic editing. Our method follows the inversion-free difference-flow paradigm, extends it to a unified audio-video trajectory, and introduces an adaptive spectral trajectory preservation mechanism for more consistent cross-modal editing.

\section{Method}

\begin{figure*}[t]
    \centering
    \includegraphics[width=\textwidth]{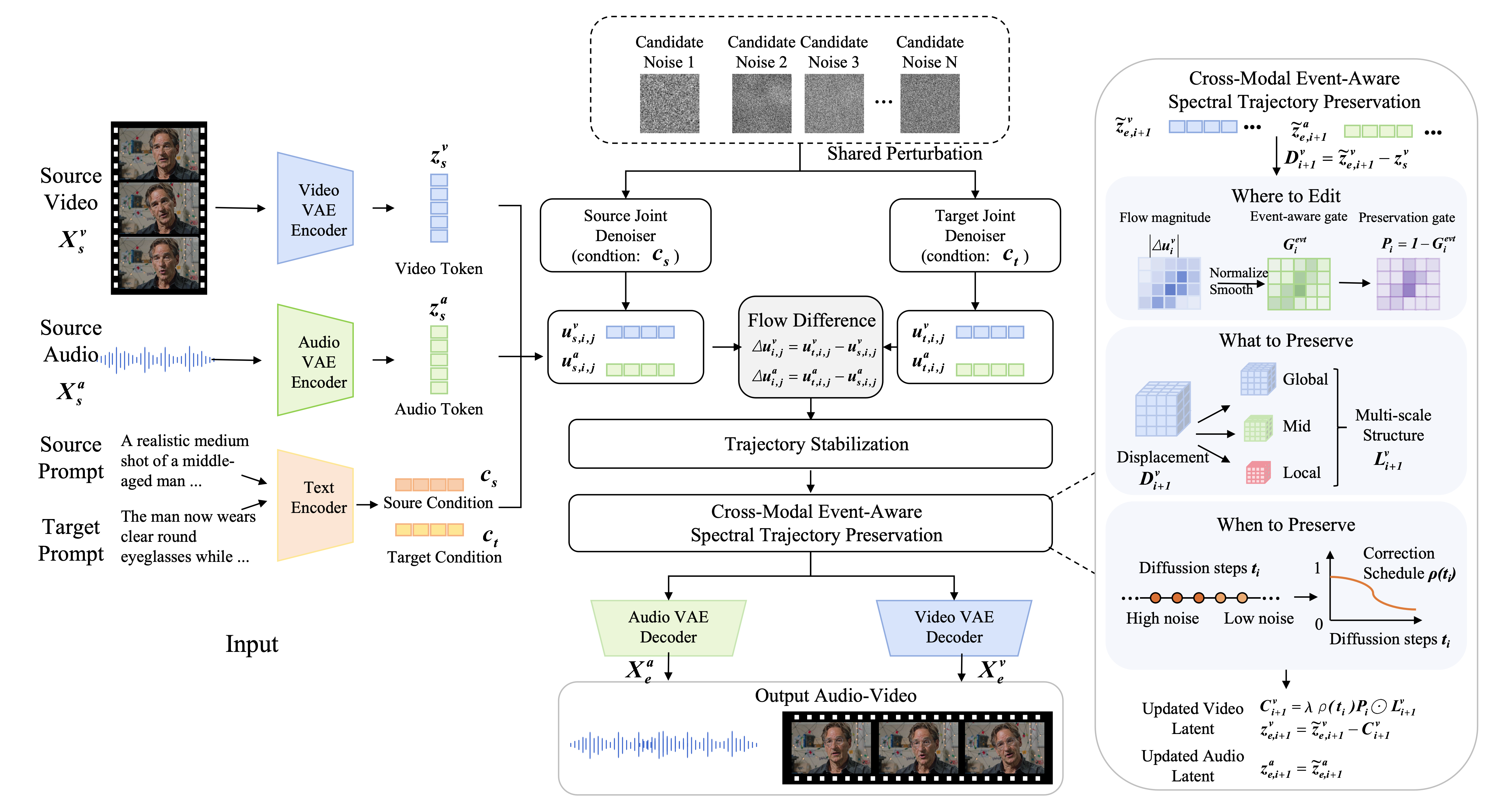}
\caption{Overall AV-FlowEdit framework. Modality-specific encoders produce separate video and audio latents, while frozen dual branches exchange cross-modal information to estimate coupled source--target editing directions. Flow saliency, multiscale spatiotemporal decomposition, and diffusion scheduling jointly guide spectral trajectory preservation.}    \label{fig:framework}
\end{figure*}

\subsection{Problem Formulation and Overall Framework}

Given a source audio-video clip $\mathbf{X}_{\mathrm{s}}=(\mathbf{X}_{\mathrm{s}}^{\mathrm{v}},\mathbf{X}_{\mathrm{s}}^{\mathrm{a}})$, its source prompt, and a target editing prompt, our goal is to generate an edited clip $\mathbf{X}_{\mathrm{e}}=(\mathbf{X}_{\mathrm{e}}^{\mathrm{v}},\mathbf{X}_{\mathrm{e}}^{\mathrm{a}})$. The output should follow the target prompt while preserving source content outside the intended edit and maintaining temporal correspondence between visual and acoustic events.

As shown in Fig.~\ref{fig:framework}, our framework addresses two coupled challenges in training-free audio-video editing. First, it must obtain reliable cross-modal editing directions without additional training or diffusion inversion. To this end, we encode audio and video into modality-specific latents and estimate source--target difference flows with a frozen joint audio-video denoiser, followed by trajectory stabilization. Second, it must preserve non-target structure without suppressing the intended edit. We address this challenge with Cross-Modal Event-Aware Spectral Trajectory Preservation, which combines event-aware flow saliency, multiscale structural decomposition, and diffusion-stage scheduling.

\subsection{Modality-Specific Encoding and Joint Difference Flow}

\subsubsection{Video, Audio, and Text Representations}

We use LTX-2~\cite{hacohen2026ltx} as the backbone and retain its modality-specific representation spaces. The pretrained video and audio VAEs, together with their token-packing operations, encode the source clip as

\begin{equation}
\mathbf{z}_{\mathrm{s}}^{\mathrm{v}}
=\mathcal{E}_{\mathrm{v}}(\mathbf{X}_{\mathrm{s}}^{\mathrm{v}}),
\qquad
\mathbf{z}_{\mathrm{s}}^{\mathrm{a}}
=\mathcal{E}_{\mathrm{a}}(\mathbf{X}_{\mathrm{s}}^{\mathrm{a}}).
\label{eq:representations}
\end{equation}

Here, $\mathcal{E}_{\mathrm{v}}$ and $\mathcal{E}_{\mathrm{a}}$ denote the frozen video and audio encoders, respectively. In our notation, $\mathbf{X}$ denotes an observation in the original video or audio space, $\mathbf{z}$ denotes its packed latent representation, the subscript $\mathrm{s}$ identifies the source clip, and the superscripts $\mathrm{v}$ and $\mathrm{a}$ identify the video and audio modalities. Thus, $\mathbf{z}_{\mathrm{s}}^{\mathrm{v}}$ and $\mathbf{z}_{\mathrm{s}}^{\mathrm{a}}$ are the source video and audio latents, respectively; the modality superscripts are labels rather than exponents. The frozen text encoder maps the source and target prompts to the conditions $\mathbf{c}_{\mathrm{s}}$ and $\mathbf{c}_{\mathrm{t}}$ shown in Fig.~\ref{fig:framework}. The packed video latent retains its spatiotemporal token coordinates, while the audio latent follows the native temporal organization of the audio branch. Video and audio tokens remain in their respective frozen branches and exchange information through cross-attention. We initialize each edited latent from its source counterpart:

\begin{equation}
\mathbf{z}_{\mathrm{e},0}^{m}=\mathbf{z}_{\mathrm{s}}^{m},
\qquad m\in\{\mathrm{v},\mathrm{a}\},
\label{eq:edit_initialization}
\end{equation}

Here, the subscript $\mathrm{e}$ denotes the edited trajectory, while $m\in\{\mathrm{v},\mathrm{a}\}$ specifies the video or audio modality. The index $0$ marks the initial trajectory state before the first editing update.

\subsubsection{Joint Editing Direction under Shared Perturbations}

Following the inversion-free FlowEdit formulation~\cite{kulikov2025flowedit}, we derive the editing direction from the difference between source- and target-conditioned velocity fields. Let
\[
t_0>t_1>\cdots>t_N=0
\]
denote the decreasing flow-time schedule, where $t_i$ is the flow time at the $i$-th sampling step.

At step $i$, we use $K_i$ candidate perturbation pairs,
\[
\left\{
\left(
\boldsymbol{\epsilon}_{i,j}^{\mathrm{v}},
\boldsymbol{\epsilon}_{i,j}^{\mathrm{a}}
\right)
\right\}_{j=1}^{K_i},
\]
for the video and audio modalities. Here, $K_i$ is the number of candidates at step $i$, and $j$ indexes one candidate pair. Each candidate contains a video perturbation $\boldsymbol{\epsilon}_{i,j}^{\mathrm{v}}$ and an audio perturbation $\boldsymbol{\epsilon}_{i,j}^{\mathrm{a}}$. The perturbations follow standard Gaussian distributions and are progressively correlated across sampling steps, as described in the next subsection.

For each candidate $j$ and modality $m\in\{\mathrm{v},\mathrm{a}\}$, we construct paired source- and target-conditioned states as

\begin{equation}
\begin{aligned}
\mathbf{x}_{\mathrm{s},i,j}^{m}
&=(1-t_i)\mathbf{z}_{\mathrm{s}}^{m}
+t_i\boldsymbol{\epsilon}_{i,j}^{m},\\
\mathbf{x}_{\mathrm{t},i,j}^{m}
&=\mathbf{z}_{\mathrm{e},i}^{m}
+\mathbf{x}_{\mathrm{s},i,j}^{m}
-\mathbf{z}_{\mathrm{s}}^{m}.
\end{aligned}
\label{eq:paired_states}
\end{equation}

The source state $\mathbf{x}_{\mathrm{s},i,j}^{m}$ interpolates between the source latent $\mathbf{z}_{\mathrm{s}}^{m}$ and the candidate perturbation $\boldsymbol{\epsilon}_{i,j}^{m}$ at flow time $t_i$. The target state transfers the same perturbation-induced offset,
$\mathbf{x}_{\mathrm{s},i,j}^{m}-\mathbf{z}_{\mathrm{s}}^{m}$,
to the current edited latent $\mathbf{z}_{\mathrm{e},i}^{m}$. Thus, within each modality, the source- and target-conditioned predictions are evaluated under the same perturbation.

For $\ell\in\{\mathrm{s},\mathrm{t}\}$, let
\[
\mathbf{x}_{\ell,i,j}
=
\left(
\mathbf{x}_{\ell,i,j}^{\mathrm{v}},
\mathbf{x}_{\ell,i,j}^{\mathrm{a}}
\right)
\]
collect the video and audio states. The frozen joint audio-video denoiser
$\mathcal{F}_{\boldsymbol{\theta}}$, implemented with separate video and audio branches, predicts

\begin{equation}
\mathbf{u}_{\ell,i,j}
=
\mathcal{F}_{\boldsymbol{\theta}}
\left(
\mathbf{x}_{\ell,i,j};
\mathbf{c}_{\ell},t_i
\right),
\qquad
\ell\in\{\mathrm{s},\mathrm{t}\},
\label{eq:conditioned_flows}
\end{equation}

where
\[
\mathbf{u}_{\ell,i,j}
=
\left(
\mathbf{u}_{\ell,i,j}^{\mathrm{v}},
\mathbf{u}_{\ell,i,j}^{\mathrm{a}}
\right)
\]
contains the predicted video and audio velocities, and
$\mathbf{c}_{\mathrm{s}}$ and $\mathbf{c}_{\mathrm{t}}$ are the source and target text conditions, respectively.

The candidate editing direction for modality $m$ is defined as

\begin{equation}
\Delta\mathbf{u}_{i,j}^{m}
=
\mathbf{u}_{\mathrm{t},i,j}^{m}
-
\mathbf{u}_{\mathrm{s},i,j}^{m}.
\label{eq:difference_flow}
\end{equation}

Because the source and target predictions use the same perturbation, their difference primarily captures the change induced by the target condition and the current editing trajectory. Moreover, each forward pass jointly processes the video and audio branches, allowing
$\Delta\mathbf{u}_{i,j}^{\mathrm{v}}$ and
$\Delta\mathbf{u}_{i,j}^{\mathrm{a}}$
to retain the cross-modal interactions of the pretrained model.

\subsection{Trajectory Stabilization}

Following DynaEdit~\cite{kulikov2026versatile}, we stabilize the difference flow using cross-step correlated perturbations and source-consistent candidate aggregation. For candidate $j$, the perturbation evolves as

\begin{equation}
\boldsymbol{\epsilon}_{i,j}^{m}
=\alpha_i\boldsymbol{\epsilon}_{i-1,j}^{m}
+\sqrt{1-\alpha_i^2}\,\boldsymbol{\eta}_{i,j}^{m},
\qquad
\boldsymbol{\eta}_{i,j}^{m}\sim\mathcal{N}(\mathbf{0},\mathbf{I}),
\label{eq:correlated_noise}
\end{equation}

where $\alpha_i\in[0,1]$ controls cross-step correlation and $\boldsymbol{\epsilon}_{0,j}^{m}$ is initialized from the same Gaussian distribution. We project each candidate direction to the clean endpoint and aggregate candidates according to their source similarity:

\begin{equation}
\begin{aligned}
\widehat{\mathbf{z}}_{i,j}^{m}
&=\mathbf{z}_{\mathrm{e},i}^{m}-t_i\Delta\mathbf{u}_{i,j}^{m},\\
w_{i,j}^{m}
&=\operatorname{softmax}_{j}\!\left(
\operatorname{sim}(\widehat{\mathbf{z}}_{i,j}^{m},
\mathbf{z}_{\mathrm{s}}^{m})/\tau\right),\\
\Delta\mathbf{u}_{i}^{m}
&=\frac{\mathbf{z}_{\mathrm{e},i}^{m}
-\sum_{j=1}^{K_i}w_{i,j}^{m}\widehat{\mathbf{z}}_{i,j}^{m}}{t_i}.
\end{aligned}
\label{eq:stabilized_flow}
\end{equation}

Here, $\operatorname{sim}(\cdot,\cdot)$ is mean-centered cosine similarity and $\tau>0$ is the softmax temperature. With $h_i=t_i-t_{i+1}>0$ and modality-specific edit scale $\gamma_m$, the temporary next state is

\begin{equation}
\widetilde{\mathbf{z}}_{\mathrm{e},i+1}^{m}
=\mathbf{z}_{\mathrm{e},i}^{m}
-h_i\gamma_m\Delta\mathbf{u}_{i}^{m}.
\label{eq:temporary_update}
\end{equation}

This state is subsequently corrected by the preservation module.

\begin{table*}[t]
\centering
\caption{Quantitative comparison of different audio-video editing methods.}
\label{tab:main_results}
\resizebox{\linewidth}{!}{
\begin{tabular}{c|c|cccccccc}
\toprule
Video Model & Audio Model 
& CLIP-F$\uparrow$
& CLIP-T$\uparrow$
& Obj.$\uparrow$
& DINO$\uparrow$
& CLAP$\uparrow$
& LPAPS$\downarrow$
& IB.$\uparrow$
& AV-Align$\uparrow$\\
\midrule

ControlVideo & SDEdit 
&0.883&0.255&0.176&0.892&0.190&6.93&0.21&0.29\\

TokenFlow & SDEdit
&0.876&0.252&0.173&0.924&0.190&6.93&0.19&0.26\\

RAVE & SDEdit
&0.885&0.251&0.170&0.881&0.190&6.93&0.18&0.31\\

ControlVideo & ZEUS
&0.883&0.255&0.176&0.892&0.211&6.41&0.21&0.30\\

TokenFlow & ZEUS
&0.876&0.252&0.173&0.924&0.211&6.41&0.20&0.27\\

RAVE & ZEUS
&0.885&0.251&0.170&0.881&0.211&6.41&0.18&0.31\\

RAVE & Diff-Foley
&0.885&0.251&0.170&0.881&0.191&7.33&0.16&0.35\\

\multicolumn{2}{c|}{DDS}
&0.890&0.250&0.175&0.921&0.210&5.93&0.20&0.33\\

\multicolumn{2}{c|}{AvED}
&0.903&0.260&0.180&\textbf{0.956}&0.226&5.55&0.23&\textbf{0.42}\\

\midrule

\multicolumn{2}{c|}{\textbf{AV-FlowEdit (Ours)}}
&\textbf{0.920}&\textbf{0.273}&\textbf{0.673}&0.592
&\textbf{0.255}&\textbf{4.10}&\textbf{0.320}&0.297\\

\bottomrule
\end{tabular}
}
\end{table*}

\begin{figure*}[t]
    \centering
    \includegraphics[width=\textwidth]{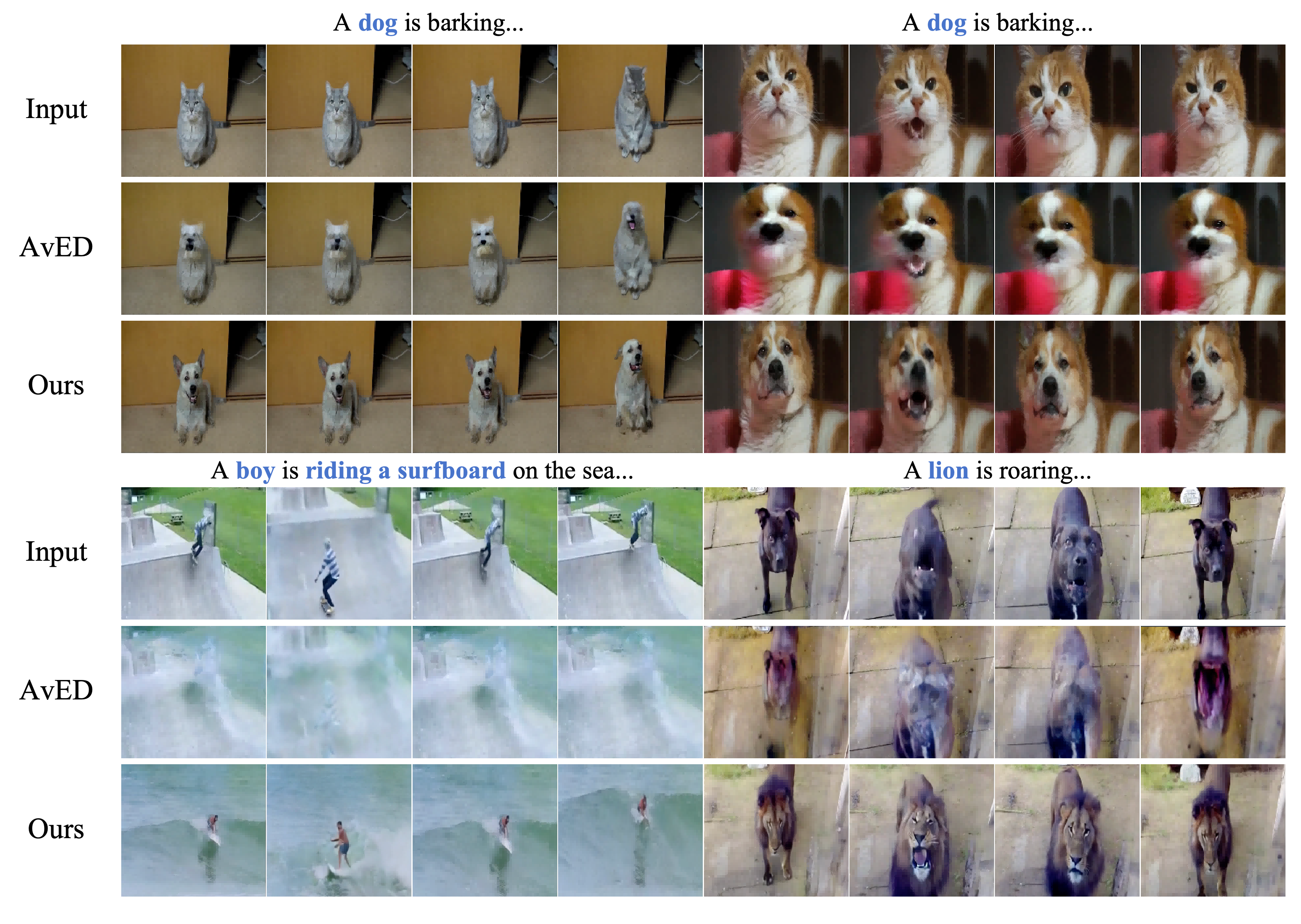}
    \caption{Qualitative comparison with AvED on prompt-driven audio-video editing.}
\label{fig:qualitative}
\end{figure*}

\subsection{Cross-Modal Event-Aware Spectral Trajectory Preservation}

Although the stabilized difference flow drives the requested edit, repeated updates can accumulate unintended displacement in the background, subject geometry, and camera motion. We therefore correct the temporary video state in Eq.~\eqref{eq:temporary_update} by determining \textit{where} editing is supported (and thus where to preserve), \textit{what} structural displacement should be removed, and \textit{when} the correction should be applied.

\subsubsection{Flow Saliency for Determining Where to Edit}

We use the magnitude of the stabilized video difference flow as a self-derived edit signal. For video token $n$, its instantaneous saliency is

\begin{equation}
s_{i,n}^{\mathrm{v}}
=\left\|\Delta\mathbf{u}_{i,n}^{\mathrm{v}}\right\|_2,
\qquad
\overline{s}_{i,n}^{\mathrm{v}}
=\beta_{\mathrm{s}}\overline{s}_{i-1,n}^{\mathrm{v}}
 +(1-\beta_{\mathrm{s}})s_{i,n}^{\mathrm{v}},
\label{eq:flow_saliency}
\end{equation}

where $\beta_{\mathrm{s}}\in[0,1)$ is the cross-step EMA coefficient. We apply quantile normalization and spatiotemporal smoothing to $\overline{\mathbf{s}}_{i}^{\mathrm{v}}$. To incorporate event timing, we also aggregate the magnitude of $\Delta\mathbf{u}_{i}^{\mathrm{a}}$ over audio time bins and align the resulting activity profile to video-token times. Modulating the smoothed visual saliency with this aligned audio activity produces the event-aware edit gate $\mathbf{G}_{i}^{\mathrm{evt}}\in[0,1]^{N_{\mathrm{v}}}$ shown in Fig.~\ref{fig:framework}, where $N_{\mathrm{v}}$ is the number of packed video tokens. Without an audio edit signal, the gate is determined by visual flow saliency alone. Large gate values indicate locations supported by the target-conditioned flow difference. The corresponding preservation gate is

\begin{equation}
\mathbf{P}_{i}=\mathbf{1}-\mathbf{G}_{i}^{\mathrm{evt}},
\label{eq:event_gate}
\end{equation}

where $\mathbf{1}\in\mathbb{R}^{N_{\mathrm{v}}}$ is the all-ones vector. Visually salient tokens are released more strongly when they coincide with the aligned audio event, whereas weakly supported tokens receive larger preservation weights.

\subsubsection{Spatiotemporal Multiscale Decomposition for Determining What to Preserve}

We measure the accumulated video displacement after the temporary flow update as

\begin{equation}
\mathbf{D}_{i+1}^{\mathrm{v}}
=\widetilde{\mathbf{z}}_{\mathrm{e},i+1}^{\mathrm{v}}
 -\mathbf{z}_{\mathrm{s}}^{\mathrm{v}}.
\label{eq:video_displacement}
\end{equation}

Let $\mathcal{R}$ map the packed video tokens to their dense spatiotemporal grid in $\mathbb{R}^{d_{\mathrm{v}}\times T\times H\times W}$, where $d_{\mathrm{v}}$ is the video-latent channel dimension and $(T,H,W)$ are the temporal, height, and width dimensions of the video-token grid. For each scale $r\in\mathcal{K}=\{\mathrm{local},\mathrm{mid},\mathrm{global}\}$, a three-dimensional low-pass operator $\mathcal{L}_{r}$ extracts structural displacement at a different receptive field. The combined low-frequency displacement is

\begin{equation}
\mathbf{L}_{i+1}^{\mathrm{v}}
=\mathcal{R}^{-1}\!\left[
  \sum_{r\in\mathcal{K}}
  \pi_r\mathcal{L}_{r}
  \!\left(\mathcal{R}(\mathbf{D}_{i+1}^{\mathrm{v}})\right)
\right],
\qquad
\pi_r\geq0,
\quad
\sum_{r\in\mathcal{K}}\pi_r=1.
\label{eq:multiscale_displacement}
\end{equation}

Small receptive fields capture short-range structural deviations, while larger fields capture object-level motion and global scene-layout drift. Combining these scales avoids tying preservation to a single spatiotemporal frequency range.

\subsubsection{Diffusion Schedule for Determining When to Preserve}

The flow schedule controls when structural correction is most active. With $t_i\in[0,1]$ decreasing from the high-noise to the clean endpoint, we use

\begin{equation}
p(t_i)=t_i^{\kappa},
\qquad \kappa>0,
\label{eq:preservation_schedule}
\end{equation}

which applies stronger correction while coarse scene structure is formed and gradually relaxes it toward the low-noise endpoint. The final correction and next video state are

\begin{equation}
\begin{aligned}
\mathbf{C}_{i+1}^{\mathrm{v}}
&=\lambda p(t_i)
  \mathbf{P}_{i}\odot\mathbf{L}_{i+1}^{\mathrm{v}},\\
\mathbf{z}_{\mathrm{e},i+1}^{\mathrm{v}}
&=\widetilde{\mathbf{z}}_{\mathrm{e},i+1}^{\mathrm{v}}
  -\mathbf{C}_{i+1}^{\mathrm{v}},\\
\mathbf{z}_{\mathrm{e},i+1}^{\mathrm{a}}
&=\widetilde{\mathbf{z}}_{\mathrm{e},i+1}^{\mathrm{a}},
\end{aligned}
\label{eq:preservation_correction}
\end{equation}

where $\lambda\geq0$ is the overall preservation strength and $\odot$ denotes token-wise multiplication with broadcasting over latent channels. The audio branch is not directly corrected; instead, its temporally aligned activity is absorbed into $\mathbf{G}_{i}^{\mathrm{evt}}$ before the preservation gate is formed in Eq.~\eqref{eq:event_gate}. After the final step, the frozen VAE decoders reconstruct the edited modalities:

\begin{equation}
\mathbf{X}_{\mathrm{e}}^{\mathrm{v}}
=\mathcal{D}_{\mathrm{v}}(\mathbf{z}_{\mathrm{e},N}^{\mathrm{v}}),
\qquad
\mathbf{X}_{\mathrm{e}}^{\mathrm{a}}
=\mathcal{D}_{\mathrm{a}}(\mathbf{z}_{\mathrm{e},N}^{\mathrm{a}}),
\label{eq:decoding}
\end{equation}

where $\mathcal{D}_{\mathrm{v}}$ and $\mathcal{D}_{\mathrm{a}}$ include inverse token packing and the video and audio VAE decoders, respectively. The decoded streams are synchronized using the source frame rate and duration to form $\mathbf{X}_{\mathrm{e}}$.












\section{Experiments}
\subsection{Baselines}

We compare AV-FlowEdit with representative visual editing, cascaded audio-visual editing, and joint audio-visual editing methods.

For independent editing pipelines, we select ControlVideo~\cite{zhang2024controlvideo}, TokenFlow~\cite{geyer2024tokenflow}, and RAVE~\cite{kara2024rave} as video editing models, and combine them with audio editing models SDEdit~\cite{meng2021sdedit} and ZEUS~\cite{manor2024zero} to establish audio-visual editing baselines.

For cascaded editing pipelines, we adopt RAVE together with Diff-Foley~\cite{luo2023diff}, where the edited video serves as the condition for subsequent audio generation.

For joint audio-visual editing, we compare with Delta Denoising Score (DDS)~\cite{hertz2023delta} and Zero-shot Audio-Visual Editing via Cross-modal Delta Denoising (AvED)~\cite{lin2026zero}.

\subsection{Evaluation Metrics}

To comprehensively evaluate AV-FlowEdit for audio-visual joint editing, 
we assess the performance from three perspectives: video editing quality, 
audio editing quality, and cross-modal consistency.

\textbf{Video Quality and Editing Consistency.}
We adopt CLIP-F, CLIP-T, Obj., and DINO to evaluate video editing quality.
CLIP-F measures temporal semantic consistency across frames, CLIP-T evaluates 
the alignment between edited videos and target prompts, Obj. measures target 
object generation accuracy, and DINO evaluates visual structure preservation.

\textbf{Audio Quality and Editing Consistency.}
We use CLAP and LPAPS to evaluate audio editing quality.
CLAP measures audio-text semantic alignment, while LPAPS evaluates perceptual 
consistency between the source and edited audio.

\textbf{Audio-Visual Consistency.}
ImageBind similarity (IB.) and AV-Align are used to evaluate cross-modal consistency, 
measuring high-level audio-visual semantic alignment and temporal synchronization 
between visual events and audio events.

\subsection{Experimental Results}

Tab.~\ref{tab:main_results} presents quantitative comparisons between AV-FlowEdit and existing audio-video editing methods in terms of video editing quality, audio editing quality, and cross-modal consistency. Overall, AV-FlowEdit achieves a better balance among target semantic editing capability, source content preservation, and audio-visual consistency under a training-free and inversion-free setting.

The results in Tab.~\ref{tab:main_results} and Fig.~\ref{fig:qualitative} demonstrate that AV-FlowEdit achieves strong performance in target semantic editing, obtaining superior results on CLIP-T and Obj. compared with existing methods. This indicates that the proposed method can effectively understand target prompts and generate content that satisfies the desired editing objectives. Meanwhile, CLIP-F and DINO reflect the inherent trade-off between semantic modification and source content preservation. Since AV-FlowEdit enables larger semantic changes, the generated results may exhibit more noticeable visual differences from the source video, leading to an impact on the DINO score, which measures source-target visual similarity. Overall, AV-FlowEdit achieves a favorable balance between editing flexibility and content preservation.

For audio editing and cross-modal consistency, AV-FlowEdit achieves competitive performance on CLAP, LPAPS, and IB., demonstrating that joint audio-video flow trajectory modeling can effectively exploit cross-modal information to generate semantically consistent audio while maintaining coordination between visual and audio events. Compared with independent modality editing pipelines and cascaded approaches, AV-FlowEdit performs unified trajectory modeling within a frozen joint audio-video Transformer, resulting in more coherent audio-visual editing outcomes.

\begin{table}[t]
\centering
\caption{Ablation of the adaptive trajectory preservation mechanism.}
\label{tab:ablation}
\begingroup
\small
\setlength{\tabcolsep}{3.2pt}
\renewcommand{\arraystretch}{1.08}
\begin{tabular}{@{}lcccc@{}}
\toprule
\multicolumn{1}{c}{Variant}
& \makecell{CLIP-T$\uparrow$\\Video}
& \makecell{Obj.$\uparrow$\\Video}
& \makecell{CLAP$\uparrow$\\Audio}
& \makecell{AV-Align$\uparrow$\\Joint} \\
\midrule
w/o spectral
& 0.263 & 0.668 & 0.236 & 0.286 \\
Fixed
& 0.270 & 0.671 & 0.251 & 0.294 \\
Adaptive
& \textbf{0.273} & \textbf{0.673}
& \textbf{0.255} & \textbf{0.297} \\
\bottomrule
\end{tabular}
\endgroup
\end{table}

\subsection{Further Analysis}



The adaptive trajectory preservation mechanism is designed to alleviate the conflict between target semantic editing and source content preservation during the editing process. To evaluate the contribution of each component, we construct three variants:

\textbf{(1) w/o Spectral Preservation:} The trajectory preservation module is removed, and editing is performed only based on the joint source-target flow difference and trajectory stabilization mechanism.

\textbf{(2) Fixed Preservation:} A fixed-strength structural preservation constraint is introduced without considering editing saliency or diffusion-stage information.

\textbf{(3) Adaptive Preservation (Full):} The complete adaptive trajectory preservation mechanism is adopted, which dynamically adjusts the preservation strength according to editing saliency, multi-scale structural information, and diffusion scheduling.




















As shown in Tab.~\ref{tab:ablation}, removing the trajectory preservation mechanism still enables semantic editing, but leads to degraded structural stability. Introducing fixed preservation constraints improves content consistency; however, the lack of adaptive adjustment according to editing regions and diffusion stages may limit the desired semantic transformation. In contrast, Adaptive Preservation achieves a better balance between structural stability and editing capability, obtaining improved performance on CLIP-T, Obj., CLAP, and AV-Align, which validates the effectiveness of the adaptive trajectory preservation mechanism.

\section{Conclusion}

In this work, we present AV-FlowEdit, a unified, training-free, and inversion-free framework for prompt-driven joint audio-video editing. Experimental results demonstrate that AV-FlowEdit outperforms existing training-free editing methods in terms of spatial consistency, temporal coherence, and audio-visual semantic alignment. This work validates the effectiveness of unified audio-video editing based on joint flow trajectory modeling and provides a promising direction for future controllable multimodal generation and editing.

\bibliographystyle{aaai}
\bibliography{references}

\end{document}


\title{AV-FlowEdit: Training-Free Prompt-Driven Audio-Visual Editing with Event-Aware Spectral Trajectory Preservation\\Supplementary Material}
\maketitle

\section{Supplementary Overview}
This supplementary material provides additional implementation details, evaluation protocols, qualitative results, user study details, efficiency analysis, and limitations.

\begin{figure}[t]
    \centering
    \includegraphics[width=0.8\columnwidth]{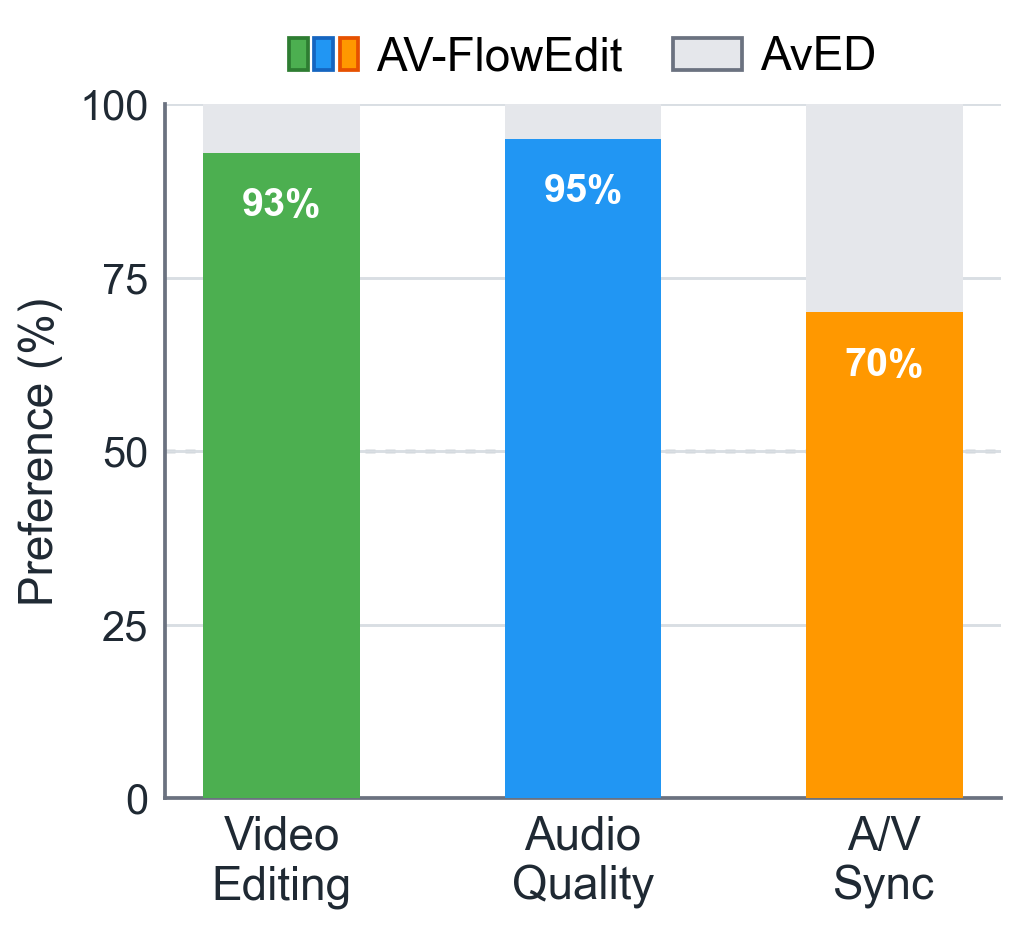}
    \caption{
        Results of the blinded pairwise user study comparing AV-FlowEdit
        with AvED.
    }
    \label{fig:user-study-preference}
\end{figure}

\begin{table}[t]
\centering
\caption{Efficiency on one A800-SXM 80 GB GPU for a 121-frame, $640\times384$ clip. Report at least 10 runs as mean $\pm$ standard deviation.}
\label{tab:efficiency}
\small
\setlength{\tabcolsep}{3.6pt}
\begin{tabular}{@{}lccc@{}}
\toprule
Method & Time (min)$\downarrow$ & Memory (GB)$\downarrow$ & Calls \\
\midrule
Single-candidate core & $6.21\!\pm\!0.10$ & $42.08\!\pm\!0.00$ & 76 \\
AV-FlowEdit & $9.14\!\pm\!0.03$ & $42.08\!\pm\!0.00$ & 124 \\
AvED & $11.61\!\pm\!0.10$ & $22.08\!\pm\!0.04$ & 2,994 \\
\bottomrule
\end{tabular}
\end{table}

\section{Implementation Details}
\label{sec:implementation}

\subsection{Backbone and Preprocessing}

We use the frozen LTX-2.3 22B checkpoint without additional training.
All model components remain frozen.

Input videos are processed into 121 frames at 12 FPS and resized to
$640\times384$.
Audio is resampled to 16 kHz and aligned with the video duration.
Experiments are conducted on one NVIDIA A800-SXM GPU with 80 GB memory.
The complete inference configuration is provided in Table~\ref{tab:hyperparameters}.Table~\ref{tab:efficiency} summarizes the inference efficiency of AV-FlowEdit and the compared methods under the same experimental setting.

\section{Benchmark Construction}
The main evaluation contains 93 source clips with source and target descriptions. Clips are drawn from VGGSound-derived audio-video material used by the AvED benchmark. Each row specifies source and target prompts and categories.

\section{User Study}
\label{sec:user_study}

\paragraph{Design}
We conducted a blinded pairwise user study comparing AV-FlowEdit with AvED. Ten participants each evaluated ten source videos, yielding 100 pairwise judgments for each criterion. In each trial, participants were shown the source clip, the editing prompts, and two edited clips with audio. The left--right order of the two methods was randomized, and their identities were hidden. Participants were allowed to replay the clips and completed a headphone playback check before the study. For each trial, they independently selected the preferred result according to the following three criteria:
\begin{enumerate}
    \item \textbf{Video editing quality:} Which result better realizes the target visual edit while preserving unrelated visual content?
    \item \textbf{Audio quality:} Which result produces more natural and artifact-free audio?
    \item \textbf{Audio-video synchronization:} Which result exhibits better semantic and temporal alignment between visible and audible events?
\end{enumerate}
\paragraph{Results.}
As shown in Fig.~\ref{fig:user-study-preference}, participants consistently preferred AV-FlowEdit over AvED across all three criteria. AV-FlowEdit was selected in 93\% of the comparisons for video editing quality and in 95\% for audio quality. It was also preferred in 70\% of the comparisons for audio-video synchronization. These results indicate that AV-FlowEdit produces more effective visual and audio edits while maintaining stronger correspondence between the two modalities.
\begin{table*}[t]
\centering
\caption{Main-benchmark inference settings. Audio event weight is the contribution of aligned audio activity to the visual release gate.}
\label{tab:hyperparameters}
\small
\setlength{\tabcolsep}{5pt}
\begin{tabular}{@{}llllll@{}}
\toprule
Group & Parameter & Value & Group & Parameter & Value \\
\midrule
Sampling & Inference steps & 40 & Sampling & Random seed & 0 \\
Sampling & Source CFG & 2.5 & Sampling & Target CFG & 6.0 \\
Sampling & Video scale $\gamma_{\mathrm{v}}$ & 1.0 & Sampling & Audio scale $\gamma_{\mathrm{a}}$ & 1.0 \\
Stabilization & Candidate count & 4 & Stabilization & Candidate steps & 8 \\
Stabilization & Temperature $\tau$ & 0.05 & Stabilization & Correlation power & 1.0 \\
Stabilization & Skipped initial steps & 2 & Preservation & Strength $\lambda$ & 0.05 \\
Preservation & Saliency quantile & 0.55 & Preservation & Gate sharpness & 1.5 \\
Preservation & Saliency EMA & 0.3 & Preservation & Gate EMA & 0.85 \\
Preservation & Audio event weight & 0.0 & Preservation & Max. correction ratio & 0.08 \\
Preservation & Local kernel $(T,H,W)$ & $(3,3,3)$ & Preservation & Mid-scale kernel & $(5,5,5)$ \\
Preservation & Global kernel & $(9,9,9)$ & Preservation & Mid/global weights & $0.25/0.10$ \\
Preservation & Motion radius $(T,H,W)$ & $(1,1,1)$ & Preservation & Boundary feather & 3 \\
Preservation & Schedule & Late-only & Preservation & Masked low-pass & Enabled \\
\bottomrule
\end{tabular}
\end{table*}
\begin{figure*}[t]
    \centering
    \includegraphics[width=\textwidth]{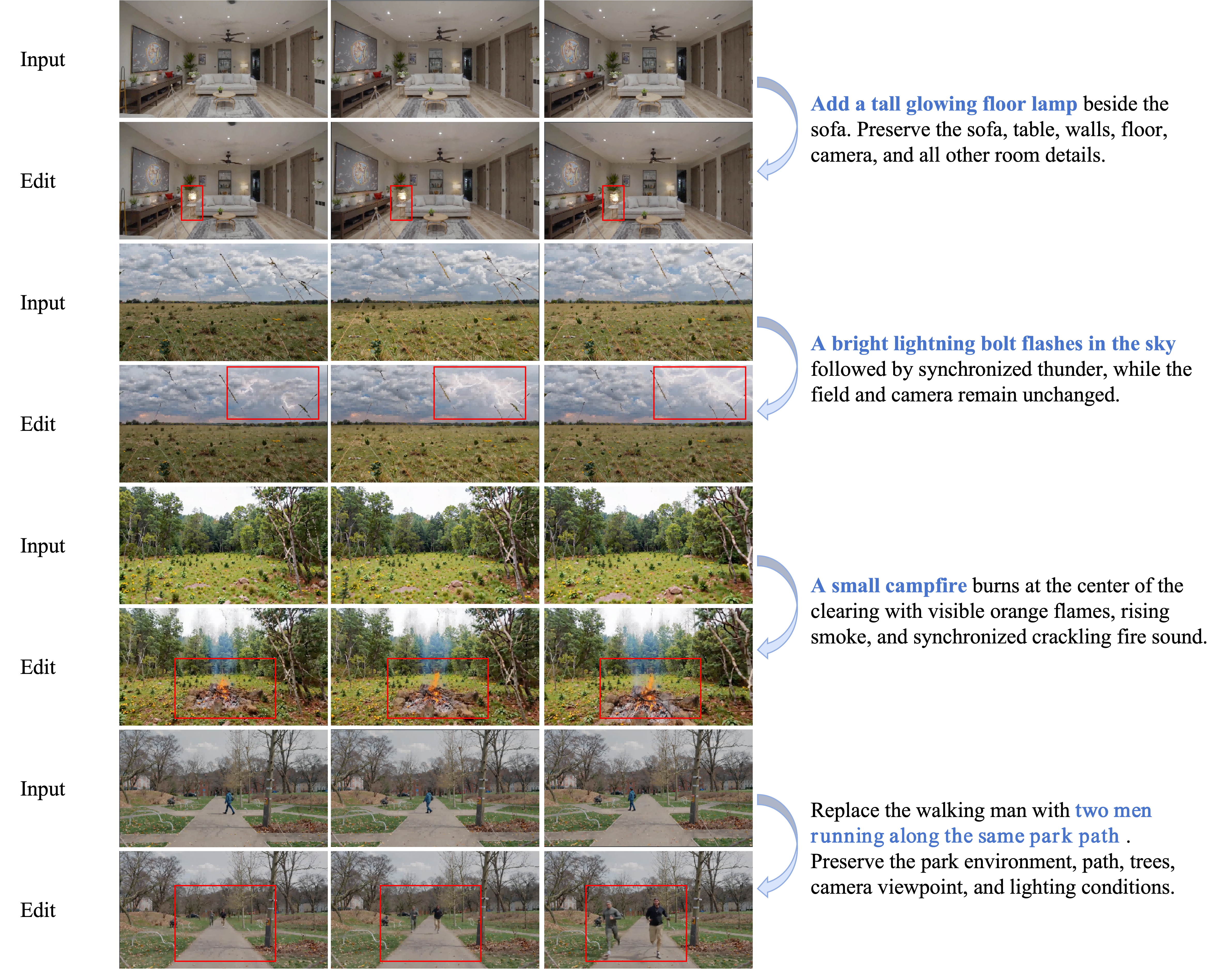}
    \caption{
Additional qualitative results of AV-FlowEdit.
The examples demonstrate diverse audio-visual editing cases, including motion changes, object modifications, and environment transformations.
AV-FlowEdit preserves irrelevant content while accurately following the target prompts.
}
    \label{fig:intro}
\end{figure*}
\bibliographystyle{aaai}